\documentclass[11pt]{article}
\usepackage[sfdefault]{carlito} % Use Carlito, similar to Calibri
\usepackage{geometry}
\usepackage{setspace}
\usepackage{titlesec}
\usepackage{hyperref}
\usepackage{graphicx}
\usepackage{amsmath}
\usepackage{bm}
\usepackage{wrapfig}
\usepackage{xcolor}
\usepackage{authblk}
\usepackage{ulem}
\usepackage[font=small,labelfont=bf]{caption}
\usepackage[symbol]{footmisc}

\titleformat{\section}{\normalfont\fontsize{11}{13}\bfseries\sffamily}{\thesection}{1em}{}
\titleformat{\subsection}{\normalfont\fontsize{11}{13}\bfseries\sffamily}{\thesubsection}{1em}{}
\titleformat{\title}{\normalfont\fontsize{14}{16}\bfseries\sffamily}{}{0em}{}

\title{Molecular Transport}

\author[1,2]{\textbf{Mar\'ia {Camarasa-G\'omez$^\ast$}}}
\affil[1]{Departamento de Polímeros y Materiales Avanzados: F\'isica, Qu\'imica y Tecnolog\'ia, Facultad de Qu\'imica, UPV/EHU, Apartado 1072, 20018 Donostia-San Sebasti\'an, Spain}
\affil[2]{Centro de F\'isica de Materiales CFM/MPC (CSIC-UPV/EHU), Paseo Manuel de Lardizabal 5, 20018 Donostia-San Sebasti\'an, Spain}

\author[3]{\textbf{Daniel {Hernang\'{o}mez-P\'{e}rez}$^\ast$}}
\affil[3]{CIC nanoGUNE BRTA, Tolosa Hiribidea 76, 20018 San Sebasti\'an, Spain}

\author[4]{Jan {Wilhelm}}

\affil[4]{Institute of Theoretical Physics and Regensburg Center for Ultrafast Nanoscopy (RUN), University of Regensburg, 93040 Regensburg, Germany}

\author[5]{\textcolor{black}{Alexej Bagrets}}
\affil[5]{\textcolor{black}{Institute of Nanotechnology, Karlsruhe Institute of Technology, 76344 Eggenstein--Leopoldshafen, Germany, and Steinbuch Centre for Computing, Karlsruhe Institute of Technology, 76344 Eggenstein-Leopoldshafen, Germany (\textsl{currently at ITK Engineering GmbH)}}}

\author[4]{\textbf{Ferdinand {Evers}}$^\ast$}

\date{}

\usepackage{seqsplit}

\begin{document}
\maketitle

\textcolor{black}{\textsuperscript{*}: Coordinator of this contribution}

\section*{Summary}

Single-molecule junctions \cite{Su2016, Evers2018, Evers2020, Cuevas2010} - nanoscale systems where a molecule is connected to metallic electrodes - offer a unique platform for studying charge, spin and energy transport in non-equilibrium many-body quantum systems, with few parallels in other areas of condensed matter physics.
Over the past decades, %
these systems have revealed a wide range of remarkable quantum phenomena, including quantum interference \cite{Su2016, Camarasa2020}, non-equilibrium spin-crossover \cite{Evers2011b}, diode-like behavior \cite{Elbing2005}, or chiral-induced spin selectivity \cite{Evers2021}, among many others \cite{Evers2018, Evers2020}.
{\color{black} To develop a detailed understanding, 
it turned out } 
essential to have {\color{black} available} \textit{ab initio}-based tools for accurately describing quantum transport in such systems \cite{Evers2018, Evers2020}. % 
{\color{black} They need to be}  capable of capturing the intricate electronic structure of molecules, sometimes in the presence of electron-electron or electron-phonon interactions, in out-of-equilibrium environments. %
Such tools {\color{black} are indispensable also for} 
experimentally observed phenomena 
{\color{black} explained in terms of parametrized} tight-binding models for the quantum transport problem.

{\color{black} While FHI-aims also offers specialized transport routines \cite{Havu2011,Havu2012, Ketolainen2017}, e.g. for chemically functionalized nanotubes or nanotube networks, our focus in this section  is on the AITRANSS package \cite{Arnold2007, Camarasa2024, Bagrets2013, Wilhelm2013} designed for simulations of single-molecule transport.}
AITRANSS is an independent post-processing tool that combined with FHI-aims enables the  calculation of electronic transport properties, as well as atom-projected density of states, spin properties and the simulation of scanning tunneling microscope images in molecular junctions. Pilot versions of the code extend some of these capabilities to non-linear transport  in the applied bias, with plans to include these features in future releases of the package.

\section*{Current Status of the Implementation}

The AITRANSS code implements the non-equilibrium Green's function formalism (NEGF) \cite{Evers2020, Cuevas2010, Jauho2008, DiVentra2008}; AITRANSS can handle closed-shell and spin-polarized contacts~\cite{Bagrets2013} and also spin-orbit coupling \cite{Camarasa2024, Camarasa2021}.
A central element in the calculation of electronic transport is the (ballistic) transmission function, which can be obtained using the trace formula \cite{Wingreen1992}
\begin{equation}\label{eq:trans}
T(E)  = \textnormal{Tr} \left[\hat{\Gamma}_L \hat{G}(E) \Gamma_R \hat{G}^\dagger (E) \right],
\end{equation}
where $\hat{G}(E)$ denotes the Green's function of the extended molecule in the presence of the leads, \textit{i.e.} source and drain. The extended molecule includes  a part of the leads as metal clusters, which are in contact with the molecule; the construction of AITRANSS offers various advantages,  \textit{e.g.}, complete flexibility in the relative orientation of the electrodes. As described elsewhere \cite{Evers2020}, the Green's function can be obtained through  partitioning, 
\begin{equation}\label{eq:green}
 \hat{G} = (E1\!\!1 - \hat{\mathcal{H}}^{\textnormal{KS}} -  \hat{\Sigma})^{-1}.
\end{equation}
Here, $1\!\!1$ represents the identity operator, $\hat{\mathcal{H}}^ {\textnormal{KS}}$ the Kohn-Sham Hamiltonian and $\hat{\Sigma} = \hat{\Sigma}_{L} + \hat{\Sigma}_{R}$ the self-energy operator of the left ($L$) and right ($R$) leads. Finally, $\hat{\Gamma}_{L,R}$ denotes the anti-Hermitian part of the self-energy operators $\Sigma_{L,R}$.
A distinctive feature of AITRANSS is its computationally efficient implementation of absorbing boundary condition using a  model self-energy approach \cite{Arnold2007, Bagrets2013, Evers2011}. In this scheme, the self-energy is given by
\begin{equation}
    \hat{\Sigma}_{\alpha} = \sum_{\tilde{\mu}, \tilde{\nu} \in \mathcal{S}_{\alpha}}|{\tilde{\mu}}\rangle[\delta \epsilon-i\eta]\delta_{\tilde{\mu}\tilde{\nu}}\langle{\tilde{\nu}}|,
\end{equation}
where $\delta \epsilon$ is the real energy shift, and $\eta$ characterizes the imaginary part, which corresponds to a local, energy-independent, material-specific leakage rate out of the scattering region. 
Note that the self-energy is applied only within a subspace $\mathcal{S}_\alpha$ corresponding to the lead atoms of the extended molecule that are farthest away from the molecule. 

\begin{figure}
    \centering
    \includegraphics[width=0.5\textwidth]{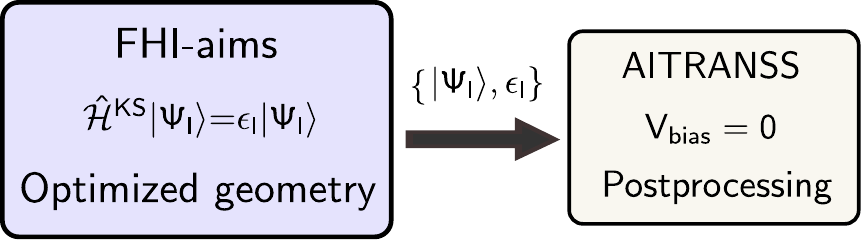}
    \caption{Workflow of the post-processing-only  cycle combining FHI-aims and the AITRANSS package.}
    \label{fig:f1}
\end{figure}

\textit{Calculations with post-processing only.} The simplest  operation mode of AITRANSS is non-self-consistent, see Fig. \ref{fig:f1}, \textit{i.e.} without feedback loop. The process begins with a standard DFT calculation using the optimized geometry of the molecular junction obtained from FHI-aims. It delivers the set of Kohn-Sham energies $\{\epsilon_l\}$ and orbitals 
\begin{equation}
 \Psi_l(\mathbf{r}) = \sum_{m  = 1}^{N_\mathcal{B}} B_{m l} \varphi_m(\mathbf{r}),
\end{equation}
where $N_\mathcal{B}$ is the number of orbitals, $B_{ml}$ are the molecular~orbital coefficients and $\varphi_m$ denotes the FHI-aims basis set. 
These orbitals are used by AITRANSS to reconstruct the Hamiltonian. Because the basis is non-orthogonal, AITRANSS employs the L\"owdin orthogonalization procedure \cite{Lowdin1950} using the overlap matrix, $S$, 
\begin{equation}\label{eq:lowdin}
 \varphi_{\tilde{m}}(\mathbf{r}) = \sum_{m' = 1}^{N_\mathcal{B}} S^{-1/2}_{m',m} \varphi_m(\mathbf{r}),
\end{equation}
to \textcolor{black}{orthogonalize the states and} reconstruct the Kohn-Sham Hamiltonian in an orthogonal basis,  $\hat{\mathcal{H}}^\textnormal{KS} = S^{1/2} B \bm{\epsilon} B^\dagger S^{1/2}$.  The resulting Hamiltonian is then used for evaluating Eqs. \eqref{eq:trans}-\eqref{eq:green}. 

\begin{figure}
    \centering
    \includegraphics[width=0.5\textwidth]{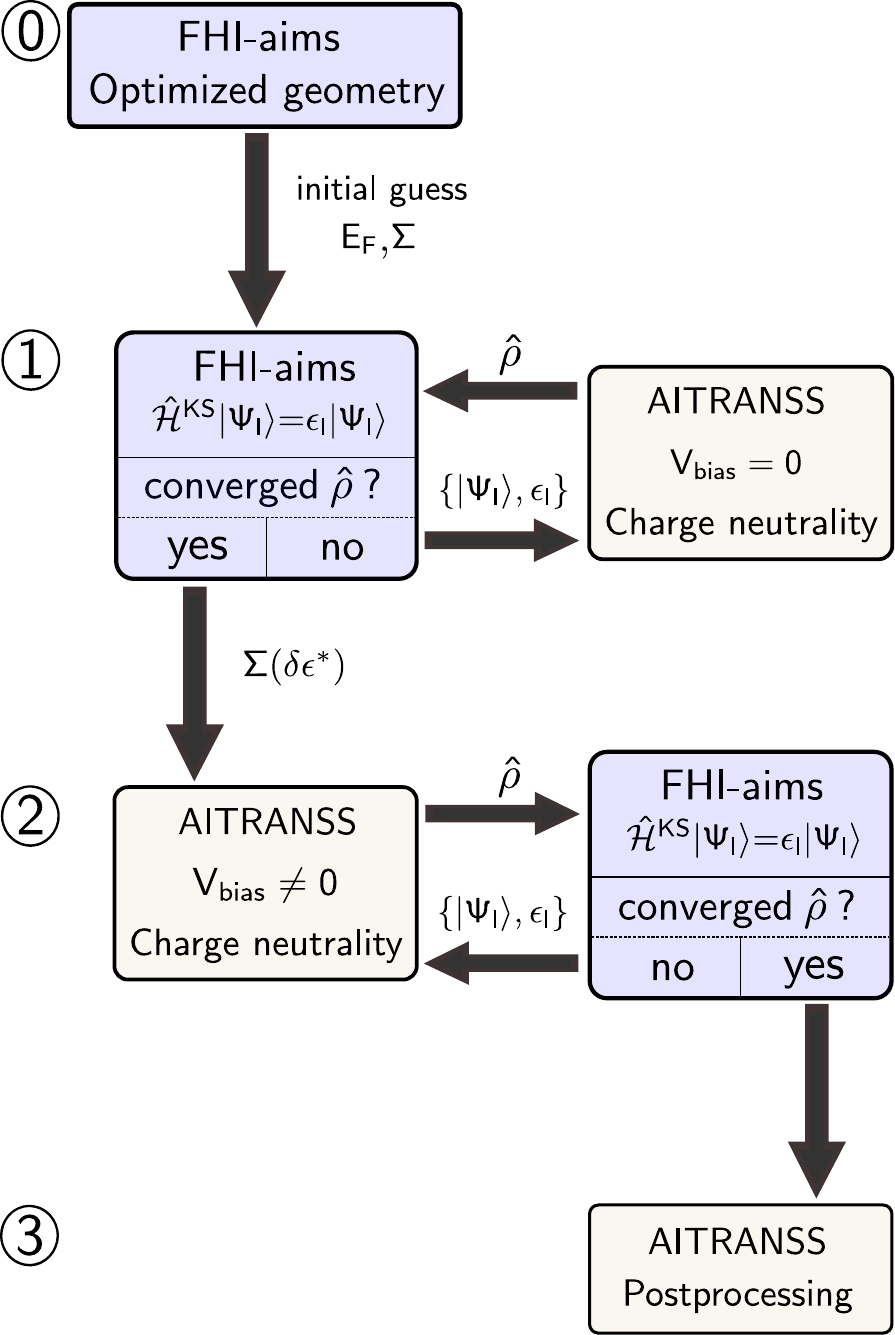}
    \caption{Workflow of the self-consistent cycle combining FHI-aims and the AITRANSS package. The calculations proceeds in four steps: \textcircled{\raisebox{-0.9pt}{0}}, initial preparatory calculation with FHI-aims for the geometry optimization; \textcircled{\raisebox{-0.9pt}{1}}, parametrization of the self-energy and the Fermi energy;  \textcircled{\raisebox{-0.9pt}{2}}, self-consistent loop at finite bias; and  \textcircled{\raisebox{-0.9pt}{3}}, postprocessing step in which observables based on the non-equilibrium density matrix can be computed. Fig. adapted from Fig. 2 in Ref. \cite{Camarasa2024} (CC BY-NC 4.0).}
    \label{fig:f2}
\end{figure}

\textit{{Self-consistent calculation procedure.}} In the self-consistent calculation mode, FHI-aims and AITRANSS perform a feedback loop, see Fig. \ref{fig:f2}. 
 As in the post-processing-only variant, the starting point is a standard DFT-calculation for the extended molecule. The role of feedback is to introduce self-consistency in the sense that the DFT-calculation in FHI-aims and the Green's function calculations in AITRANSS refer to the same (non-equilibrium) density matrix $\rho$: during each iteration step the Kohn-Sham Hamiltonian is updated with $\hat{\rho}$, which is constructed using the NEGF \cite{Camarasa2024, Bagrets2013, Camarasa2021}. The feedback-loop establishes, in particular, the Fermi energy, E\textsubscript{F}, at fixed particle number, $N$, and the real part of the self-energy, $\delta \epsilon$. Physically, the cycle ensures correct charge redistribution in the junction,  accounting for the macroscopic nature of the contacts and maintaining charge neutrality while screening the excess charge accumulated at the outermost boundaries of the finite-sized clusters \cite{Arnold2007, Camarasa2024, Bagrets2013, Camarasa2021,  Palacios2023, Naskar2023}.
During the feedback loop, the real part of the self-energy, $\Sigma(\delta \epsilon^\ast)$, is gradually deformed so as to enforce charge-homogeneity on the far sites of the metal cluster, also under finite bias voltage, V\textsubscript{bias}. Physical observables are calculated in the final post-processing step: transmission function, current-voltage characteristics, spin-orbit torques, and more.
%
%
%

%%%%%%%%%%%%%%%%%%%%%%%%%%%%%%%%%%%%%%%%%%%

\section*{Usability and Tutorials}

AITRANSS is a project under continuous development since 2002, and is currently centered at Universit\"at Regensburg. The current main developers are Mar\'ia Camarasa-G\'omez, Daniel Hernang\'omez-P\'erez and Ferdinand Evers.
The source code, along with examples and a basic electrode library, is distributed with the FHI-aims package. It can be found in the subdirectory \texttt{external/aitranss}. This directory contains the source code, the script \texttt{tcontrol.aims.x} used to prepare the mandatory \texttt{tcontrol} input file for AITRANSS, a representative electrode library, and a folder with documented examples, including input and output files of FHI-aims and AITRANSS.

Compilation instructions, explanations of the code, and a list of available keywords are provided in the FHI-aims manual.
As an independent package, AITRANSS has its own build system, which has also been integrated into the FHI-aims \texttt{cmake}-based build scheme, as documented in the FHI-aims manual release \texttt{240507}. The current release of AITRANSS is parallelized using OpenMP directives, enabling the creation of a multithreaded version of the executable.

Without spin-orbit coupling, using the mandatory keyword  \texttt{output aitranss} in the \texttt{control.in} file generates three ASCII files after a successful run of FHI-aims: \texttt{basis.out}, which contains information about the basis functions, \texttt{omat.aims}, which contains the overlap integrals and \texttt{mos.aims} which contains the Kohn-Sham orbitals and energies for the extended molecule. If an open-shell calculation is performed, \texttt{mos.aims} is substituted by two files called \texttt{alpha.aims} and \texttt{beta.aims}. AITRANSS is then run in the same directory where the output files of FHI-aims are located.
To include spin-orbit interaction, the keywords

\textcolor{black}{\texttt{include\_spin\_orbit} \\
\texttt{output soc\_eigenvectors 1} \\
\texttt{output soc\_aitranss}}

must also be used, as documented in Ref. \cite{Camarasa2021}.  %
The basis and molecular orbital files are  replaced by \texttt{basis-indices.soc.out
} and \texttt{omat.aims.soc}.
The current implementation of the self-consistent non-equilibrium cycle (Fig. \ref{fig:f2}) is managed by an external shell script and maintains its performance in terms of memory and computational requirements for small molecular junctions. Additional details can be found as well in Ref. \cite{Camarasa2021}.

Finally, as a project in continuous expansion, online tutorials and  updated information can be accessed from the AITRANSS webpage at \hyperlink{https://aitranss.ur.de/}{https://aitranss.ur.de/}.

\section*{Future Plans and Challenges}
The AITRANSS package is an ongoing project under continuous development, with new features and capabilities to be added in the near future.
From a computational perspective, we will focus on maintaining the package  efficiently. The code is currently being refactored to enhance  readability, making it more compact and easier to manage. Additional examples and tutorials will be added as well.
In addition, we plan to extend the parallelization strategies beyond OpenMP (multithreading) by incorporating MPI parallelization, which will also function independently from the former. Further enhancements include replacing outdated ASCII files with modern formats such as HDF5.

From the functionality perspective, a significant issue, resulting from the limitations of Kohn-Sham transport calculations using semilocal functionals, is the deviation of Kohn-Sham spectral properties from the exact values. We aim to address this issue by employing scissor-like operation techniques based on hybrid functionals and image-charge correction.
Furthermore, we will be expanding the study of transport and dynamical properties in the presence of spin-orbit interactions, for example incorporating spin-orbit effects in scanning tunneling microscope imaging simulations.
We also plan to continue developing the implementations for current-induced forces (both mechanical and spin) and to incorporate interaction with light in the terahertz regime, local currents \cite{Walz2014, Walz2015, Wilhelm2015} as well as phonon effects.
Finally, we plan to work on implementing multiterminal calculations in AITRANSS, particularly in the context of electrochemistry, where electrostatic effects are crucial. 

\section*{Acknowledgments}
We appreciate pleasant and fruitful discussions with many colleagues over the years: L. Venkataraman, M. S. Inkpen, R. Koryt\'ar,  M. Kamenetska, 
\textcolor{black}{M. Walz, G. Solomon, J. van Ruitenbeek.} \textcolor{black}{Contributions of P. Havu and V. Havu to the development of additional transport routines interfaced with FHI-aims are acknowledged.}
M. C.-G. acknowledges support from the Gobierno Vasco-UPV/EHU [Project no. IT1569-22]. 
D. H.-P. is grateful for funding from the Diputaci\'on Foral de Gipuzkoa through Grants 2023-FELL-000002-01, 2024-FELL-000009-01, and from the Spanish \seqsplit{MICIU/AEI/10.13039/501100011033} through Project No. PID2023-147324NA-I00.
J. W. acknowledges funding by the German Research Foundation (DFG) via the Emmy Noether Programme (Project No.~503985532). F. E. is grateful for support from DFG through the Collaborative Research Center (SFB) 1277 - Project ID 314695032 (subproject A03),
from GRK 2905 (project-ID 502572516) and from the State Major Instrumentation Program, INST 89/560-1 (project number 464531296).

% Explicit bibliography to make it easier to format in the future

\end{document}